\documentclass[aps,twocolumn,showpacs]{revtex4} 
\usepackage{graphicx}

\begin{document} 
\draft

\newcommand{\beq}{\begin{equation}} 
\newcommand{\eeq}{\end{equation}} 
\newcommand{\ben}{\begin{eqnarray}} 
\newcommand{\een}{\end{eqnarray}} 
\newcommand{\bea}{\begin{array}} 
\newcommand{\eea}{\end{array}} 
\newcommand{\om}{(\omega )} 
\newcommand{\bef}{\begin{figure}} 
\newcommand{\eef}{\end{figure}} 
\newcommand{\leg}[1]{\caption{\protect\rm{\protect\footnotesize{#1}}}}
\newcommand{\ew}[1]{\langle{#1}\rangle} 
\newcommand{\be}[1]{\mid\!{#1}\!\mid} 
\newcommand{\no}{\nonumber} 
\newcommand{\etal}{{\em et~al }} 
\newcommand{\geff}{g_{\mbox{\it{\scriptsize{eff}}}}} 
\newcommand{\da}[1]{{#1}^\dagger} 
\newcommand{\cf}{{\it cf.\/}\ } 
\newcommand{\ie}{{\it i.e.\/}\ }

\title{\center{Single photon quantum cryptography}}

\author{Alexios Beveratos, Rosa Brouri, Thierry Gacoin$^*$, Andr\'e Villing, 
Jean-Philippe Poizat and Philippe Grangier} 
\affiliation{Laboratoire Charles Fabry de l'Institut d'Optique, UMR 8501 du 
CNRS, F-91403 Orsay, France. \\ 
$^*$ Laboratoire de Physique de la Mati\`ere Condens\'ee, 
Ecole Polytechnique, F-91128 Palaiseau, France.\\ 
{\rm e-mail : alexios.beveratos@iota.u-psud.fr}}
%
%
\begin{abstract}
We report the full implementation of a quantum cryptography protocol 
using a stream of single photon pulses generated by 
a stable and efficient source operating at room temperature. 
The single photon pulses are emitted 
on demand by a single nitrogen-vacancy (NV) color center in a diamond nanocrystal. 
The quantum bit error rate is less that $4.6 \%$ and the secure bit rate is 
$9500$ bits/s. The 
overall performances of our system reaches a domain where single photons have a 
measurable advantage over an equivalent system based on attenuated light pulses.

\end{abstract}
\maketitle

Since its initial proposal in 1984 \cite{BB84} 
and first experimental demonstration in 1992 \cite{BBBSS}, 
Quantum Key Distribution (QKD) has reached maturity through 
many experimental realizations \cite{GRTZ}, 
and it is now commercially available \cite{idq}. However, 
most of the practical realizations of QKD rely on 
weak coherent pulses (WCP) which are only approximation of single 
photon pulses (SPP), 
that would be desirable in principle. 
The presence of pulses containing two photons or more in WCPs is an 
open door to information 
leakage towards an eavesdropper. In order to remain secure, the WCP 
schemes require 
to attenuate more and more the initial pulse, as the line losses become 
higher and higher, resulting in either a vanishingly low transmission 
rate - or a loss of security \cite{BLMS,L}. 
The use of an efficient source of true single photons would therefore 
considerably improve the performances of existing or future QKD schemes, 
especially as far as high-losses schemes such as satellite QKD 
\cite{BHLMNP} are considered.

In this letter we present the first complete 
realization of a quantum cryptographic key distribution based on a 
pulsed source of 
true single photons. Our very reliable source of single photon has been used 
to send a key over a distance of 
$50$ m in free-space at a rate of $9500$ secret bits per second 
including error correction and 
privacy amplification. 
Using the published criteria that warrant 
absolute secrecy of the key against any type of individual attacks 
\cite{BLMS,L}, 
we will show that our set-up reaches the region where a single photon QKD scheme 
takes a quantitative advantage over a similar system using WCP.

Single photon sources have been extensively studied in recent years 
and a great variety of 
approaches has been proposed and implemented 
\cite{MGM,BBPG,mol,nano,QD,QDelec}. 
Our single photon source is based on the fluorescence 
of a single Nitrogen-Vacancy (NV) color center \cite{diam} 
inside a diamond nanocrystal \cite{BBGPG,BKBGPG} at room temperature. 
This molecular-like system has a lifetime of $23$ ns 
when it is contained in a $40$ nm nanocrystal \cite{BBGPG}. 
Its zero-phonon line lies 
at $637$ nm and its room temperature fluorescence spectrum ranges 
from $637$ nm to $750$ nm 
\cite{GDTFWB}. 
This center is 
intrinsically photostable: no photobleaching has been observed over a 
week of continuous 
saturating irradiation of the same center. The nanocrystals are held 
by a $30$ nm thick layer of 
polymer that has been spin coated on a dielectric mirror \cite{BBGPG}. 
The mirror is initially slightly fluorescing, but this background light 
is reduced to a negligible value by hours of full power excitation 
that leads to a complete 
photobleaching of the dielectric coating, the NV center being unaffected.


\begin{figure} 
\includegraphics[scale=0.36]{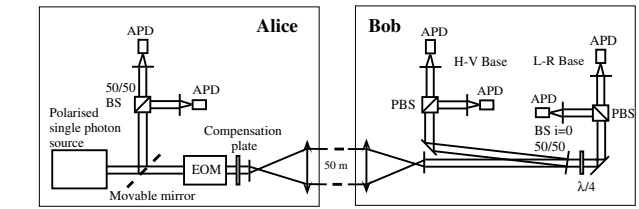} 
\caption{Experimental set-up} 
\label{expsetup} 
\end{figure}
 
The experimental set-up is shown in Fig. \ref{expsetup}. 
Alice's station consists in a pulsed single photon source, a photon 
correlation detection to 
control the quality of the SPP, and a 4-state polarization encoding scheme.
The single photon source is pumped by a home built pulsed laser at a 
wavelength of $532$ nm that delivers 
$800$ ps long pulses of energy $50$ pJ with a repetition rate of 
$5.3$ MHz, synchronized on 
a stable external clock \cite{BKBGPG}. 
The green excitation light is focused by a metallographic 
objective of high numerical aperture ($NA = 0.95$) onto the 
nanocrystals. The partially polarized fluorescence light 
(polarization rate of $46 \%$) 
is collected by the same objective. It is then polarized 
horizontally by passing through 
a polymer achromatic half-wave plate and a polarizing cube, 
spectrally filtered by a 
long-pass filter that eliminates the reflected laser light,
and spatially filtered by a confocal set-up.
In order to control the quality of the SPP, the light can be sent via 
a movable mirror onto 
a photon correlation detection scheme consisting of 
two avalanche photodiodes (APD) in a Hanbury-Brown and 
Twiss set-up.

\begin{figure} 
\includegraphics[scale=0.8]{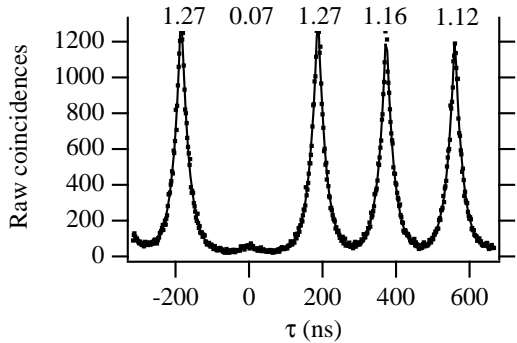} 
\caption{Autocorrelation function of a single NV center on 
Alice's side. The raw coincidences are given as a function of the 
delay between the 
arrival times of the photons at Alice's correlation detection set-up. 
The exciting laser has a repetition period of $187.5$ ns, a pulse width 
 of  $0.8$ ns and an average power of $0.2$ mW. The count rates are about 
$3.5 \times 10^4$ s$^{-1}$ on 
each avalanche photodiode, and the integration time is $166$s. The 
coincidences between peaks do not go down to zero because of the 
overlapping of adjacent peaks. The number above each 
peak represents its normalized area. The dots are experimental data. The 
line is an exponential fit for each peak and takes into account the 
background light.} 
\label{satg2} 
\end{figure}
 
The total number of polarized photons detected 
by the two APDs altogether is $N_D^{(a)}=7\times 10^4$ s$^{-1}$ 
for an excitation repetition rate of $5.3$ MHz. This gives an overall 
efficiency of $1.3 \%$. 
The autocorrelation function of the emitted light at saturation, displayed 
on Fig. \ref{satg2}, 
shows that the number of photon pairs within a pulse 
is strongly reduced with respect to Poisson statistics. 
The normalized area of the central peak is $C(0)= 0.07$, where this 
area would be 
unity for WCPs \cite{BKBGPG}. 
This means that the number of two-photon pulses of our source is 
reduced by a factor of 
$1/C(0)=14$ compared to a WCP.

The Bennett Brassard  protocol \cite{BB84} (BB84)  is implemented by using
the Horizontal-Vertical (H-V) and Circular Left-Circular 
Right (L-R) basis. These four polarization states are obtained by 
applying four levels of high voltage on an electro-optical modulator (EOM). 
The EOM is driven by a home made module, that can switch 
$500$ V in $30$ ns to ensure the $5.3$ MHz 
repetition rate of single photon source. 
The driving module is fed by sequences 
of pseudo-random numbers, that are produced using 
a linear feedback shift register in the Fibonacci configuration. 
In order to minimize polarization errors due to the 
broad bandwidth of the emitted photons, the EOM is operating very close 
to exact zero-path difference (white light fringe). This is obtained by inserting a suitable
birefringent plate  to compensate for the residual birefringence of the EOM.
Given the measured
transmission  of the EOM of $T_{\rm EOM}=0.65$, and the quantum efficiency of the 
control APDs of 
$\eta=0.6$, the rate of single photons emitted by Alice station is 
$N^{(a)}=N_D^{(a)} T_{\rm EOM}/\eta=7.58\times 10^4$ s$^{-1}$. The 
average photon number per 
pulse is thus $\mu=0.014$.

The detection at Bob's site lies $50$ m away from Alice down a corridor. The 
single photons are sent via a $2$ cm diameter beam so that 
diffraction effects are negligible. 
The H-V or L-R basis are passively selected by a near $0^o$ 
incidence $50/50$ beam splitter that is polarization insensitive. 
A polymer achromatic quarter-wave plate is inserted in the L-R basis arm. 
In each basis a polarizing beam splitter sends the two polarizations on 
two APDs. 
The time arrival of the photons is acquired by a four 
channel digital oscilloscope on a memory depth of 1 million points per 
channel and a time resolution of $10$ ns. The acquired sequence is 
hence $10$ ms long 
and it can be repeated at will after the memory of the oscilloscope 
has been emptied. 
For the sake of simplicity, the synchronization signal is send to the 
oscilloscope 
using a coaxial cable, but it would be straightforward to use the IR or green 
laser pulses for the same purpose.

The total number of photons detected by Bob is $N_D^{(b)} =3.93 
\times 10^4$ s$^{-1}$. 
The dark counts on Bob's APDs with no signal at the input are 
$(d_H,d_V,d_L,d_R)=(150,180,380,160)$ s$^{-1}$. 
This includes APD's dark counts and background noise 
due to ambient light, that is carefully shielded using dark screens 
and pinholes. Considering 
the $23$ ns lifetime of the NV center, a post selection of pulses 
within a $50$ ns gate 
selects approximatively $\eta_g=90\%$ of all single photons, and 
keeps only $\beta_g = 27\%$ 
of the background counts. After basis reconciliation, the raw bit rate 
is then $N_r = \eta_g N_D^{(b)} /2 = 1.77 \times 10^4$ s$^{-1}$. 
Taking into account the detection gate, the fraction of 
dark counts versus useful photons during 
a detection gate is therefore 
$p_{\rm dark} = \beta_g\sum_{i=H,V,L,R} d_i /(\eta_g N_D^{(b)}) = 0.7 \%$. 
The static polarization error rates are measured while 
Alice codes each one of the four polarizations, and they are 
$p_{\rm pol}^{\rm HV}=1.2 \%$ in the H-V basis 
and $p_{\rm pol}^{\rm LR} = 3.2 \%$ in the L-R basis, 
owing to the slight imperfection brought by the achromatic wave plate 
(dark counts have been substracted in these values). 
One can thus estimate the quantum bit error rate (QBER) to be 
$e = (p_{\rm dark} + p_{\rm pol}^{\rm HV} + p_{\rm pol}^{\rm LR})/2 = 2.6 \%$. 
By comparing the full 
key that Bob received to the one that Alice sent, the measured QBER 
is found to be
$e=4.6 \% \pm 1 \%$. The difference with the previous value is attributed to the fact 
that static polarization errors underestimate the real dynamic errors, 
owing to the non ideal shape of the electric pulses driving the EOM. 
The complete secret key transmission was achieved by 
carrying out error correction and privacy amplification using 
the public domain sofware ``QUCRYPT'' designed by Louis Salvail 
\cite{NSSSDP}. This leads to an average of $100$ secret bits shared by Alice and Bob 
in a $10$ ms sequence.


We now compare the performance of our single photon BB84 set-up with 
QKD schemes using WCPs \cite{BHLMNP,SGGRZ}. 
The comparison is carried out by 
taking the detection efficiency and the dark counts of Bob in the 
present set-up. 
For WCP we assume a detection gate of 
$2$ ns that is typical for recent experiments \cite{BHLMNP,SGGRZ}. 
The quantities 
that are compared are the maximum allowed on-line losses and the 
secret bit rate. 
Since QKD is supposed to offer unconditional security, it is assumed 
that a potentiel 
eavesdropper (Eve) has an unlimited technological power 
to carry out individual attacks within the rules of quantum mechanics. 
Eve can then access all the information leakage 
caused by the quantum bit error rate $e$ and by the 
multiphoton pulses \cite{BLMS}. In the case of WCP with an average 
number $\mu$ of photons 
per pulse at Alice's station ($\mu \ll 1$), 
the probability of a multiphoton pulse is given by $S_m^{\rm WCP}=\mu ^2/2$. 
The only way to reduce the fraction of multiphoton pulses in WCP is 
therefore to 
reduce the bit rate by working with smaller $\mu$. 
To the contrary, SPP offers the possibility of achieving a vanishing ratio of 
multiphoton pulses without any trade-off on the filling of the pulses. 
In the present experiment the probability of a multiphoton pulse is reduced to 
$S_m^{\rm SPP}= C(0) \mu ^2/2$ with $C(0)=0.07$.

\begin{figure} 
\includegraphics[scale=0.8]{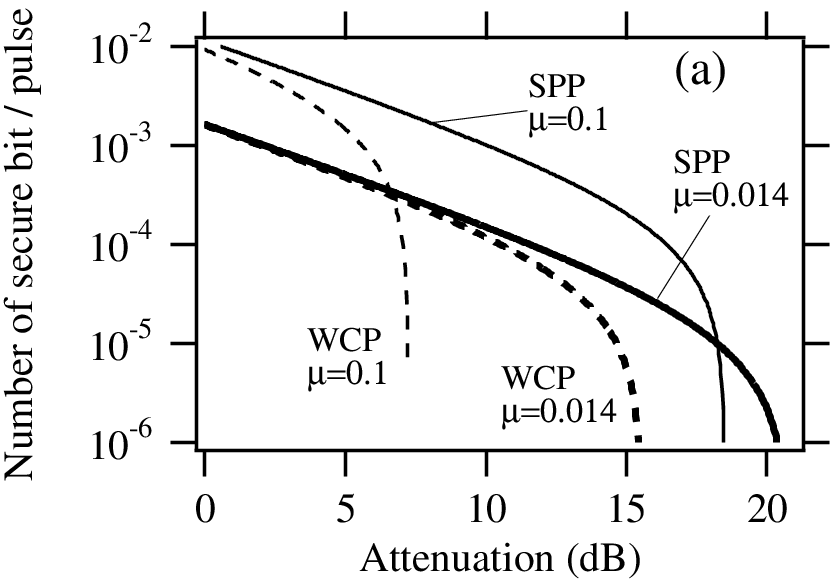} 
\vskip 0.5cm 
\includegraphics[scale=0.8]{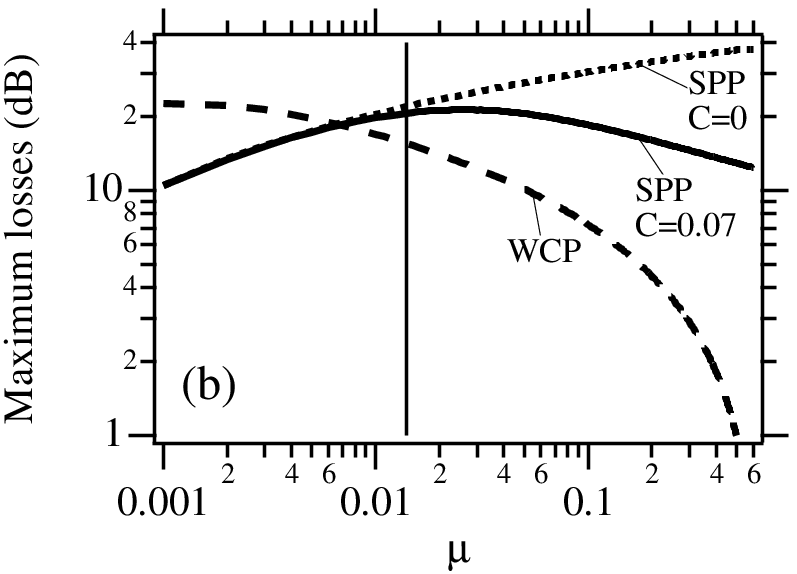} 
\vskip 0.5cm 
\includegraphics[scale=0.8]{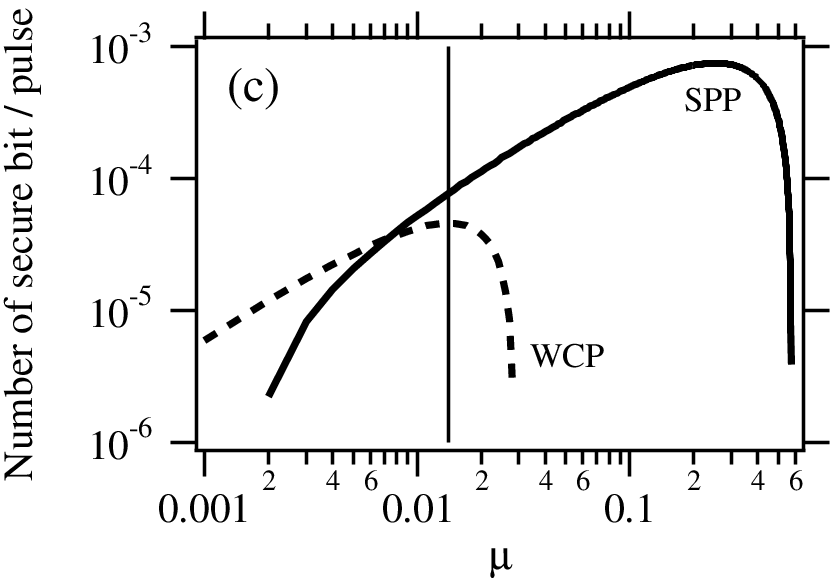} 
\vskip 0.5cm 
\caption {These plots give theoretical evaluations obtained 
by using Eq. (\ref{gainrate}), together with the 
experimental parameters for our single-photon source.
(a) Calculated number of secure bit 
per time detection time gate $G$ as a function of the on-line losses 
for SPPs and WCPs, for different average photon number per pulse $\mu$. 
The SPP traces correspond to our value of the zero time autocorrelation of $C=0.07$.
(b) The maximum allowed on-line losses for secure 
communication is deduced from (a) and corresponds 
to the attenuation for which $G=10^{-6}$. This value is plotted 
as a function of the mean 
photon number per pulse $\mu$, for a WCP system, for a SPP system with our value 
of the zero time autocorrelation ($C=0.07$), 
and for an ideal SPP system with $C=0$. The vertical line corresponds to our 
source, ie $\mu=0.014$. 
(c) Number of secure bit per detection time gate $G$ as a function of 
the mean photon number per pulse $\mu$ for on-line losses of $12.5$ dB. 
The SPP trace assumes that $C=0.07$, and 
the vertical line corresponds to $\mu=0.014$.} 

\label{gmulti} 
\end{figure}

The important figure to be evaluated is the number 
of secure bits per pulse ($G$) after error correction and privacy amplification. 
This quantity is given by \cite{L} 
\begin{eqnarray} 
\label{gainrate} 
\lefteqn{G= \frac{1}{2} p_{\rm exp} 
\bigg\{\frac{p_{\rm 
exp}-S_m}{p_{\rm exp}}}& & \nonumber \\ 
& & \times 
\left(1-\log_2 \left[1 + 4 e 
\frac{p_{\rm exp}}{p_{\rm exp}-S_m}-4\left(e 
\frac{p_{\rm exp}}{p_{\rm exp}-S_m}\right)^2\right]\right) \nonumber \\ 
& & + f[e] \left[e\log_2 
e + \left(1-e\right) \log_2 
\left(1-e\right)\right]\bigg\} \; . 
\end{eqnarray} 
The quantity $p_{\rm exp}$ is the probability that Bob has a click on 
his detectors 
(including possible dark counts) during a detection gate, and $S_m$ is
the probability of a multiphoton photon pulse just at the output of AliceÕs station.
The function $f[e]$ 
depends on the algorithm that is used for the error correction. The 
Shannon limit 
gives $f[e]=1$ for any $e$, which is the value taken in Fig. 
\ref{gmulti}. For the best known 
algorithm, 
$f[e]=1.16$ for $e\leq 5\%$. 
In our set-up, the parameters are $(p_{\rm exp},S_m,e)=(7.4\times 
10^{-3}, 1.9\times 10^{-6}, 
4.6\times 10^{-2})$ so that $G = 1.8\times 10^{-3}$. The number of secure bits 
per second is given by eq (\ref{gainrate}) is thus $N_{\rm QKD}=0.95 
\times 10^4$ s$^{-1}$ which 
is compatible with 
our experimental value of $10^4$ s$^{-1}$.

As can be seen in Fig. \ref{gmulti}, our SPP quantum cryptographic 
system has a quantitative 
advantage over the best existing WCP systems. When any type of 
individual attacks, without any technological limitations, are taken 
into account, our SPP 
system can deliver absolutely secure secret key at higher bit rate and 
offers the possibility of transmitting this key over longer distances. 
Our quantum cryptographic set-up compares also favorably with QKD 
experiments using pairs of entangled photons \cite{paires}, with a 
significantly higher secure bit rate in our case. 
Moreover, several relatively simple 
improvements could give SPP-QKD protocols an even 
greater advantage. In particular, inserting the emitter 
in a microcavity \cite{GSGCT} is within experimental reach, 
and may be helpful to increase the collection efficiency, and therefore 
the secret bit rate, and also to narrow the emission spectrum, 
and thus to reduce polarisation errors.

As a conclusion, we have demonstrated the first complete single 
photon quantum key distribution 
set-up by using a very reliable room temperature single photon source. 
Despite the 
fairly broad spectrum of the single photons, a 4-states polarization 
encoding and decoding was implemented with low error rate ($4.6\%$), 
and transmission over $50$ m in air was successfully achieved 
with a secure bit rate of $N_{\rm QKD}=9500$ s$^{-1}$. 
These results show that single photon QKD is a realistic candidate 
for long distance quantum 
cryptography, such as surface-to-satellite QKD.

We thank John Rarity for very fruitful discussions, 
Louis Salvail and Martial Tarizzo for helping us with the ``QUCRYPT'' 
sofware \cite{NSSSDP}, 
Herv\'e Rigneault for providing us with the 
back reflecting mirror, Alain Aide and Fr\'ed\'eric Moron for the 
electronics, Robert Pansu for 
the loan of the time to amplitude converter, St\'ephane Esnouf for 
the diamond irradiation and 
Marie-Fran\c coise Ravet for the diamond annealing. 
This work is part of the ``S4P'' project supported by the European 
Union IST/ FET/ QIPC program.

 

\end{document}